\documentclass[iop]{emulateapj}

\usepackage{natbib}
\usepackage{graphicx}
\usepackage{amssymb}

\usepackage{url}
\usepackage{hyperref}
 \hypersetup{
     final=true,
     pageanchor=true,
     colorlinks=true,
     breaklinks=true,
     linkcolor=blue,
     citecolor=blue,
     urlcolor=blue,
     pdfpagemode=UseNone,
     pdftitle={Magnetic and dynamical photospheric disturbances observed during an M3.2 solar flare},
     pdfauthor={C. Kuckein, M. Collados and R. Manso Sainz},
     pdfsubject={Solar Physics},
     pdfkeywords={Sun: chromosphere, Sun: flares, Sun: magnetic fields, Sun: photosphere, techniques: polarimetric}}

\shorttitle{Magnetic and dynamical photospheric disturbances observed during an M3.2 solar flare}
\shortauthors{Kuckein et al.}

\begin{document}

\title{Magnetic and dynamical photospheric disturbances observed during an M3.2 solar flare}

\author{C. Kuckein}
\affil{Leibniz-Institut f\"ur Astrophysik Potsdam (AIP), An der Sternwarte 16, 14482, Potsdam, Germany}
\email{ckuckein@aip.de}
\and

\author{M. Collados and R. Manso Sainz}
\affil{Instituto de Astrof\'\i sica de Canarias (IAC), V\'\i a L\'{a}ctea s/n, 38205, La Laguna, Tenerife, Spain}
\affil{Departamento de Astrof\'\i sica, Universidad de La Laguna, 38206 La Laguna, Tenerife, Spain}

\begin{abstract}
This letter reports on a set of full-Stokes spectropolarimetric observations in the near infrared He\,{\sc i} 10830~\AA\ 
spectral region covering the pre-, flare, and post-flare phases of an M3.2 class solar flare. The flare 
originated on 2013 May 17 and belonged to active region NOAA 11748. We detected strong He\,{\sc i}~10830~\AA\ emission 
in the flare. The red component of the He \,{\sc i} triplet peaks at an intensity ratio to the continuum
of about 1.86. During the flare, He\,{\sc i} Stokes $V$ is substantially larger and appears reversed compared
to the usually larger Si\,{\sc i} Stokes $V$ profile.  
The photospheric Si\,{\sc i} inversions of the four Stokes profiles reveal the following: (1)
the magnetic field strength in the photosphere decreases or is even absent during the flare phase, as compared
to the pre-flare phase. However, 
this decrease is not permanent. After the flare the magnetic field recovers its pre-flare configuration in a short
time (i.e., in 30 minutes after the flare). 
(2) In the photosphere, the line-of-sight velocities show a regular granular up- and down-flow pattern before the flare 
erupts. During the flare, upflows (blueshifts) dominate the area where the flare is produced. 
Evaporation rates of $\sim 10^{-3}$ and $\sim 10^{-4}$~g\,cm$^{-2}$\,s$^{-1}$ have been derived in the deep and 
high photosphere, respectively, capable of increasing the chromospheric density by a factor of two in about 400 
seconds.

\end{abstract}



\section{Introduction}

Understanding the magnetic field configuration in dynamic solar events such as flares is an unsolved problem of solar physics.
Flares are hard-to-predict events and are the result of a global change 
in the magnetic field topology. These changes mainly happen in the chromosphere and corona. Therefore, 
full-Stokes polarimetric observations of flares, especially including the chromosphere, are essential. They are, however, challenging. 
Many previous studies have focused on the photospheric magnetic field underneath flare events. 
For instance, variations
of the longitudinal magnetic field of the order of $\sim 100$~G were observed by \citet[][]{kosovichev99}, and up to almost $300$~G by \citet[][]{sudol05} below X-class flares.
An enhancement of $400$~G in the magnetic field strength was also detected by \citet[][]{kondrashova13} during the onset of a microflare.

It seems clear that changes happen in the photospheric magnetic field during flares. 
However, it is controversial whether these
variations are either associated with an increase or a decrease of the magnetic field strength. \citet[][]{petrie10} studied the photospheric magnetic field in 77 M- or X-class flares. Both, an increase and
decrease of the magnetic field was detected in different flares. These authors found absolute changes of up to $\sim 450$~G
in the longitudinal magnetic field and linked an increase of magnetic field strength to the presence of larger horizontal
fields, as observed in their work concerning flares near the limb. In agreement with this, \citet[][]{hudson08} also predicted changes leading to more horizontal fields. Recently, \citet[][]{wang12} obtained first evidence of a 
rapidly increasing horizontal magnetic field of $\sim 500$~G along the polarity inversion line (PIL)
in only 30 minutes (starting at the beginning of the flare). 
The authors conjecture that reconnection processes after the flare could trigger the formation
of low-lying horizontal fields. 

Here, we report on a set of full-Stokes spectropolarimetric observations in the near infrared in a spectral interval covering the photospheric Si\,{\sc i} 10827 \AA\ line and the chromospheric He \,{\sc i} 10830~\AA\ triplet. Using a scanning spectrograph, several maps of an active region were obtained before, during and after the development of an M3.2 flare, making it possible to derive the magnetic and dynamical disturbances at photospheric layers. Interestingly, some studies on flares have recently  been published using data obtained in the same spectral region \citep[e.g.,][]{akimov14,judge14,sasso14,zeng14}. With a data set with similar characteristics, \citet{judge14} concentrated their study on the seismic implications of the flare. These authors derived an increase of the photospheric magnetic field when comparing the maps before and after the flare. Unfortunately, they did not analyze the photospheric Stokes spectra of their map taken during the flare. \citet{sasso14} performed one scan of an active region after the onset of a flare and measured the Stokes profiles of the Si\,{\sc i} 10827 \AA\ line and of the He\,{\sc i} 10830 \AA\ triplet, deriving the magnetic structure of a filament present in the active region. Other works \citep[e.g.,][]{akimov14,zeng14} only used either He \,{\sc i} 10830~\AA\ chromospheric filtergrams or intensity spectra. In this paper, we complement previous studies by analyzing the photospheric magnetic and dynamic changes in the emission region before, during and after the flare.


\section{Observations}

\begin{figure*}[!th]
\centering
\includegraphics[width=11cm]{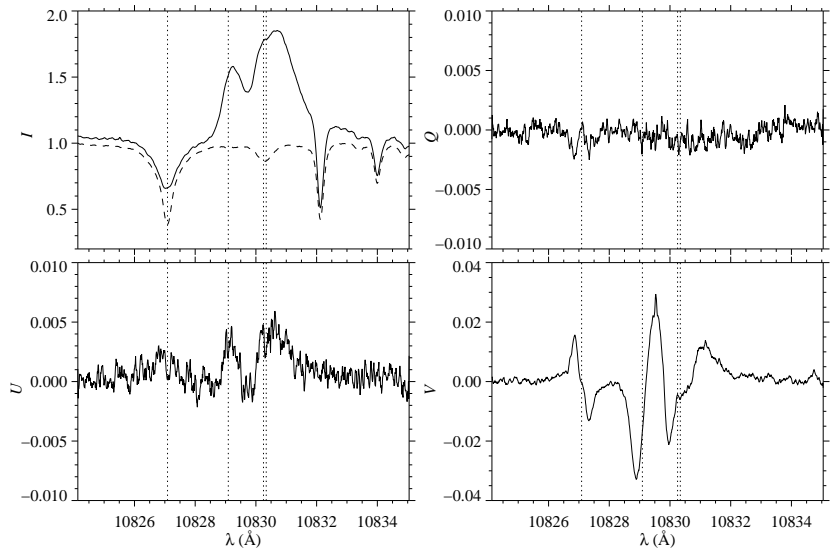}
   \caption{Example of the four normalized Stokes profiles recorded very close to the maximum of the 
flare at 8:53~UT. The exact location of the profiles is shown with an arrow in Figure \ref{fig:maps}. 
From left to right the following spectral lines are seen: Si\,{\sc i} line, He\,{\sc i} triplet, and
two telluric lines. The vertical dashed lines mark the wavelength at rest of the Si\,{\sc i} line and the He\,{\sc i}
triplet, respectively. The average quiet-Sun intensity spectrum is also plotted in dashed lines for comparison.}
\label{fig:stokesprofs}
\end{figure*}

Observations of the active region (AR) NOAA 11748 were carried out 
at the Vacuum Tower Telescope \citep[VTT, Tenerife,][]{vonder98} on 2013 May 17. 
The setup included the Tenerife Infrared Polarimeter \citep[TIP-II;][]{tip2} 
for full Stokes spectropolarimetry in the He \,{\sc i} 10830~\AA\ spectral region. 

NOAA 11748 sharply increased its activity towards 2013 May 13, when 14 flares, 
classes between C and X, occurred.
The number of flares and their activity gradually decreased in the next days. 
However, a strong M3.2 class flare was observed on 2013 May 17 while we were 
scanning with the slit along the PIL at coordinates (N~11$^\circ$, E~36$^\circ$). 
The flare started at 8:43~UT, peaked at around 8:57~UT, and ended at 9:19~UT. 
The first spatial scan was carried out between 7:48 -- 8:36~UT (pre-flare phase), the
second one between 8:36 -- 9:06~UT (flare phase), and the third one between 
9:06 -- 9:37~UT (post-flare phase), 
thus covering the whole flare activation and relaxation process.
An additional  scan of the same area was performed 
immediately after the third scan, between 9:38 -- 10:08~UT.

The exposure time per slit position was 10~s and the scanning step 0\farcs35. 
The pixel size along the slit was 0\farcs17. 
The Kiepenheuer-Institute Adaptive Optics System \citep[KAOS;][]{berkefeld10} was locked on photospheric high-contrast
structures, like pores and small penumbrae, and was crucial to improve the fair-quality seeing conditions. 
The observed spectral range with TIP-II spanned from 10824 to 10835~\AA, with a spectral sampling of
$\sim 11.1$~m\AA\,px$^{-1}$. This spectral region comprises the photospheric Si\,{\sc i} 10827~\AA\ line, the
chromospheric He\,{\sc i} 10830~\AA\ triplet, and two telluric lines.


\section{Data analysis}

Dark current, flat-field corrections and the standard polarimetric calibration for the TIP-II instrument were 
carried out \citep[][]{collados99,collados03}. 
The continuum was corrected for varying intensity due to changing air mass and 
center-to-limb variations at different positions on the solar disk while performing the scans.
This was performed by second-order least-square polynomial fits and averages over quiet-Sun 
areas on the maps. 
All pixels along the slit within one scan step were then normalized to their 
corresponding continuum value. 
Wavelength calibration is based on the two nearby telluric lines 
following the procedure described in Appendices A and B of \citet[][]{kuckein12b},
which compares the wavelength separation between telluric lines in a quiet-Sun area
with the Fourier Transform Spectrometer spectrum \citep[FTS;][]{neckel84} to obtain
the spectral sampling. 
Finally, wavelengths were corrected for Earth's orbital motions, solar rotation, and the 
solar gravity redshift \citep[][]{martinez97,kuckein12b}, thus
the line-of-sight (LOS) velocities refer to an absolute scale. 

The photospheric Si\,{\sc i} line was inverted using the Stokes Inversions
based on Response functions code  \citep[SIR;][]{SIR}. 
SIR solves the radiative transfer equation under the assumption of local thermodynamical 
equilibrium (LTE) and hydrostatic equilibrium. 
In this Letter, we focus on the inferred magnetic
field strength and the LOS velocity patterns.

The original spatial resolution was preserved, with a pixel size of
$0.35 \times 0.17$~arcsec$^2$. 
The inversions were carried out using the Harvard-Smithsonian reference atmosphere 
model \citep[HSRA;][]{gingerich71} as an initial guess for the atmosphere.  
The model atmosphere covers a range given as the logarithm of the LOS continuum optical depth $\tau$ 
at 5000~\AA\ of $1.4 \leq \log \tau \leq -4.0$. 
\citet[][]{kuckein12a} made an extensive analysis of inversion tests using different initial model atmospheres. The output atmospheres were not much dependent on the starting atmosphere, but an initial magnetic field strength of 500 G seemed to be an optimum choice, since it returned the best fits while minimizing the number of iterations required for the convergence. For this reason, we decided to add this value to the initial HSRA guess model.
The stray-light profile in each map was computed averaging the Stokes $I$ profiles 
of pixels without polarization signals. 
No more than four nodes in depth were
necessary for any of the free parameters to achieve good fits to the four Stokes profiles.


\section{Results}

\subsection{Intensity and polarization profiles during the flare}

The He \,{\sc i} 10830~\AA\ triplet originates between the lower term 
$2^3\mathrm{S}_1$ and the excited term $2^3\mathrm{P}_{2,1,0}$ of orthohelium. 
It comprises three spectral lines which are called the ``blue'' component at 10829.09~\AA\ 
($2^3\mathrm{S}_1 \rightarrow 2^3\mathrm{P}_0$), 
and the (blended) ``red'' component at 10830.30~\AA\
($2^3\mathrm{S}_1 \rightarrow 2^3\mathrm{P}_{1, 2}$).

Photospheric temperature enhancements in flares have been reported by, e.g., \citet[][]{chornogor08} and 
\citet[][]{kondrashova13} using spectral lines in the visible.
Likewise, \citet[][]{xu04} detected a near-infrared intensity enhancement of 
the continuum inside an X10 white-light flare. During our M3.2 class flare, the He\,{\sc i} Stokes $I$ profile
showed a very strong emission, the red component peaking 
at an intensity of $\sim 1.86$, as seen in Figure~\ref{fig:stokesprofs}. 
Though He\,{\sc i}~10830~\AA\ emission against
the solar disk in flares has been previously seen by several authors, 
such large values have not been reported yet. 
Previous studies showed intensity ratios to the continuum of $\sim 1.36$ for a M2.0 flare 
\citep[][]{li07}; 
$\lesssim$1.30 in the decay phase of a C9.7 flare \citep[Fig. 1;][]{penn95}; 
$\lesssim$1.15 during a C2.0 flare \citep[Fig. 2;][]{sasso11}. 
These works did not show any associated  
He\,{\sc i} Stokes $Q$, $U$, and $V$ profiles.
 
\begin{figure*}[!t]
\centering
\includegraphics[width=13.3cm]{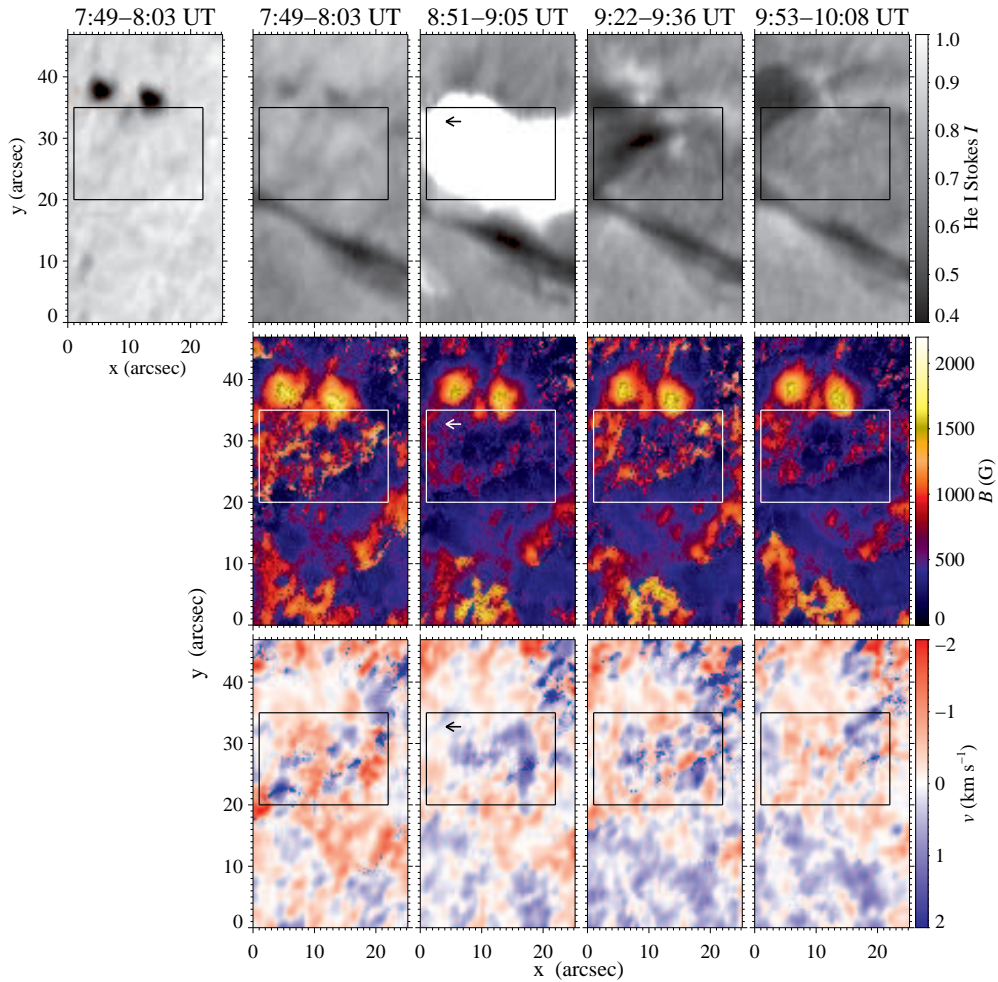}
   \caption{The isolated upper lefthand frame shows a continuum slit-reconstructed image of the 
  first map (pre-flare). Next, from {top} to {bottom}: He\,{\sc i} monochromatic red-component 
   slit-reconstructed images saturated at 
    an intensity of $I = 1$ (line emission); 
    total photospheric magnetic field strength $B$ (in Gauss) and LOS Doppler velocities 
    inferred from the Si\,{\sc i} inversions (clipped between $\pm 2$ km\,s$^{-1}$). 
    From {left} to {right}: the pre-flare, flare, and two post-flare phases are shown
    with the scanning times at the top.
    The boxes outline the region-of-interest where the flare is seen, as traced by the He\,{\sc i} emission 
    seen in the 8:51--9:05~UT panel. 
    The arrow marks the position of the Stokes profiles shown in Figure~\ref{fig:stokesprofs}.}
\label{fig:maps}
\end{figure*}

Figure \ref{fig:stokesprofs} shows an illustrative example of the Stokes 
profiles that corresponds to the maximum of the flare
(the exact position is marked by an arrow in Figure~\ref{fig:maps}). 
Remarkably, the amplitude of the circular polarization Stokes $V$ profile 
from He\,{\sc i} is larger than the one from Si\,{\sc i}. 
Furthermore, since the linear polarization signals (Stokes $Q$ and $U$) in Si~{\sc i}
are also very small, 
the magnetic field is expected to be significantly stronger in the chromosphere than in the photosphere.
Note that the amplitude of the blue component of the He I Stokes $V$ profile is significantly larger than the one of 
the red component. Typically both profiles have similar amplitudes or even the blue component has a smaller amplitude. 
In addition, while the $V$-profile of the blue component looks normal (i.e, rather antisymmetric), that of the red 
component seems distorted, especially its red lobe. This might indicate that the Stokes $V$ shown in the figure is a 
combination of at least two emission profiles.

The observed reversal of the Stokes $V$ profiles of He\,{\sc i} with respect to 
Si\,{\sc i} is due to 
the He\,{\sc i} spectral line appearing in emission, whereas the Si\,{\sc i} line 
appears in absorption. 
In fact, the Si\,{\sc i} line never goes into emission in our data series, although 
it becomes noticeably shallower during the flare, 
as demonstrated by the comparison with the average quiet-Sun spectrum plotted in Figure~\ref{fig:stokesprofs}. 
This is the result of the heating of photospheric layers produced by the flare. 

The components of the He\,{\sc i} triplet appear 
redshifted with respect to their wavelength positions at rest 
(vertical dotted lines in Figure~\ref{fig:stokesprofs}),  
suggesting that at the flare peak,
the plasma is slightly moving downwards towards the photosphere. 
Two peaks in Stokes $I$ can be distinguished in the usually blended red component of He~{\sc i}, 
one of them is slightly redshifted.
During the flare, many pixels show clear linear polarization signals
in both the red and blue components of He~{\sc i} 10830~\AA\ 
(e.g., Stokes $U$ in Figure~\ref{fig:stokesprofs}), which cannot 
be easily interpreted as Zeeman or scattering signatures.
These anomalous polarization patterns will be discussed in a forthcoming paper.

\subsection{Photospheric magnetic field suppression during the different flare phases}

The He\,{\sc i} emission pattern is used here as an indicator of the flare eruption. 
To study the underlying photosphere, we concentrate on the Si\,{\sc i} line inversions. This line is 
sensitive to a range of various heights within the photosphere. 
However, we will focus on a layer corresponding to the granular height which is well represented
by $\log \tau = -1$. 

Slit-reconstructed monochromatic intensity images centered at the red component of the He\,{\sc i} 
line are presented in the first row of Figure \ref{fig:maps}.  
These maps just represent the part of the full scanned 
field of view covered by the main part of the flare emission.
From these images, it can be easily seen when the flare activates the He\,{\sc i} emission (where $I>1$). 
The second row shows the photospheric magnetic field strength, as derived from the inversion of the Si\,{\sc i} line.
We will focus on the area inside the box which corresponds to the flare maximum, as indicated by the He\,{\sc i} 
emission. In the first column, i.e., before the flare started, 
the photospheric magnetic field inside the box shows a 
patchy pattern of strong fields. However, during the flare (second column), the field strength
globally decreased being even absent in some areas. Interestingly, in the post-flare panel (third column), 
the magnetic field has partially recovered its strength and pattern. Hence, there is clear evidence
that the photospheric magnetic field changes during the flare and afterwards tries to recover its initial 
configuration. 
A further inspection of the Stokes profiles reveals that the larger magnetic
field strength in the pre- and post-flare maps is mainly due to the larger amplitude of the Stokes $V$ profiles.
The second map after the flare (fourth column) shows that the magnetic field is again losing strength and its
pattern looks more like the one which corresponded to the flare. These fluctuations may indicate that the magnetic 
field may be undergoing temporal variations with a long relaxation time of hours.

As for the Doppler velocities presented in the last row of Figure \ref{fig:maps}, a granular pattern, i.e.,
a mix of up and downflows is seen in the pre-flare panel. In response to the flare, the second panel shows a 
large area of upflows. Hence, photospheric material is predominantly pushed upwards where the flare occurs. 
The first post-flare panel still shows mainly upflows but a granular pattern is already distinguishable. 
In the last panel the granulation pattern is re-established.

\begin{table}[!h]
\begin{center}
\caption{Mean Doppler velocities $\langle v \rangle$ and mass flux $\langle \rho v 
\rangle$, at the layer of $\log \tau = -1$, inside the box of Figure~\ref{fig:maps}.\label{tab:vel}}
\begin{tabular}{ccccc}
\tableline\tableline
Time       & Phase & $\langle v \rangle$  & $ \langle \rho v \rangle$            \\
(h:m UT)   &       & (km\,s$^{-1}$)       & (g cm$^{-2}$ s$^{-1}$) \\
\tableline
7:49--8:03 & pre-flare  & $-0.16 \pm 0.02$ & $-2.2 \times 10^{-3}$ \\ 
8:51--9:05 & flare      & $+0.11 \pm 0.01$ & $+1.5 \times 10^{-3}$ \\
9:22--9:36 & post-flare & $+0.07 \pm 0.01$ & $+1.1 \times 10^{-3}$ \\
9:53--10:08& post-flare & $-0.08 \pm 0.01$ & $-1.0 \times 10^{-3}$ \\
\tableline
\end{tabular}
\tablecomments{There are 5368 pixels in each box. 
Negative (positive) values for downflows (upflows) along the LOS.}
\end{center}
\end{table}


\section{Discussion}

We have shown a very remarkable and unusual data set including pre-, flare, and post-flare slit scans in the 
infrared He\,{\sc i} 
10830\,\AA\ spectral region of an M3.2 class flare with the spectropolarimeter TIP-II . 
Our observations reveal complex Stokes $Q$, $U$, and $V$ profiles of the chromospheric He\,{\sc i} 10830 triplet during
the flare. Furthermore, unprecedented large intensity ratios to the continuum of the He\,{\sc i} red component where 
shown ($I \sim 1.86$). However, the He\,{\sc i} emission is not associated with the whole area covered by 
the flare, as already detected by previous studies \citep[][]{du08}.
From the Si\,{\sc i} inversions we have inferred the magnetic fields and the LOS velocities in the photosphere 
for the pre-, flare, and post-flare phases.
There are roughly 30 minutes between the flare and the post-flare observations.
As seen in the color scale in Figure \ref{fig:maps}, 
the magnetic field goes through changes of up to 1500~G in 30 minutes. These changes mainly happen in 
the box outlined in Figure \ref{fig:maps}, which lies between the upper two pores and the lower PIL. 
The maximum of the flare is neither located at the sunspot nor at the PIL of the active region. 
In our data sets, it was found that 90 minutes seemed to be an upper limit for the time needed by the magnetic field strength to recover most of its pre-flare pattern and strength.
However, the variations of the magnetic field are not permanent. There is a strong decrease during the flare, 
with areas where the magnetic field was present before the flare but is completely absent during the evolution of this  
phenomenon.

\begin{figure}[!t]
\epsscale{1.1}
\plotone{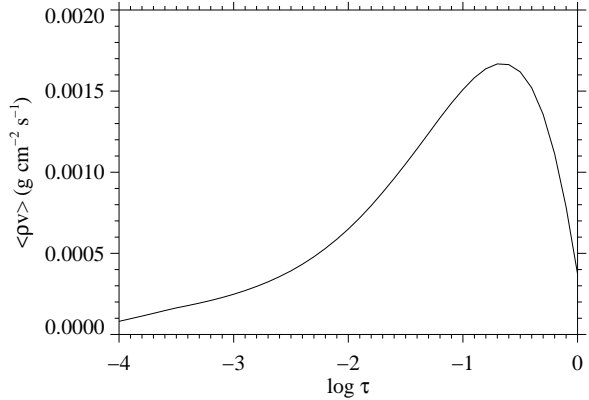}
\caption{Stratification with logarithm of the optical depth ($\log \tau$) of the average evaporation rate 
$\langle \rho v \rangle$ in the flaring region indicated by the rectangular box of the second column in 
Figure \ref{fig:maps}.}
\label{fig:massflux}
\end{figure}

Table \ref{tab:vel} shows some parameters inferred from the photospheric inversions in the flaring part of the region (represented by the box of Figure 
\ref{fig:maps}), at the layer at $\log\tau = -1$. 
Before the flare occurs, a downflow of around 160~m\,s$^{-1}$ is present in the average granular velocity.
During the flare and immediately after it, the granular pattern appears very distorted and 
barely perceptible, with an average upflow of some 100~m\,s$^{-1}$. Later, in the last map, the 
granulation pattern is almost recovered with an average downflow slightly below 100~m\,s$^{-1}$. This evolution 
is compatible with a scenario where, once the flare has started, the deep photospheric material heats up and evaporates 
at a rate of $\sim 10^{-3}$~g\,cm$^{-2}$\,s$^{-1}$.

Figure \ref{fig:massflux} shows the average dependence with depth          
of the evaporation rate of the photosphere in the flaring region.
From this stratification, the density change in the chromosphere 
can be estimated from the continuity equation:                                                                           
             
\begin{equation}                                                                                 
\frac{\partial{\rho}}{\partial t} = -\frac{d (\rho v)}{dz},                                      
\label{eq:1}                                                                                     
\end{equation}                                                                                
where we have assumed that the material motion is essentially vertical 
and the measured LOS velocity, $v$, does not differ significantly from 
the true vertical velocity. 
Integrating equation~(\ref{eq:1}) in height from the temperature minimum 
($z_\mathrm{min}$) up to the top of the atmosphere ($z_\mathrm{max}$), one gets 

\begin{equation}
\frac{\partial\sigma}{\partial t} = 
(\rho v)_{z_\mathrm{min}} - (\rho v)_{z_\mathrm{max}},
\label{eq:2}                                                                                     
\end{equation}                   
where $\sigma=\int_{z_\mathrm{min}}^{z_\mathrm{max}}\rho dz$ is the column 
mass density of the chromosphere.                                                                
Further assuming that the rate of change of the density is constant                              
during the flare and that no mass escapes away                                                   
($(\rho v)_{z_\mathrm{max}}=0$), then the time interval                                          
required to produce a change $\Delta\sigma$ in the column density is                             

\begin{equation}                                                                                 
\Delta t = \frac {\Delta \sigma} {(\rho v)_{z_\mathrm{min}}}.                                    
\label{eq:3}                                                                                     
\end{equation}                                                                                  
We can evaluate $(\rho v)_{z_\mathrm{min}}$ as the mass flux at the top                          
of our photospheric inversions (i.e., $\sim 10^{-4}$ g\,cm$^{-2}$\,s$^{-1}$;                     
see Figure \ref{fig:massflux}).                                                                  
For the aforementioned HSRA model atmosphere, the chromospheric column density                   
(i.e, integrating the density from the temperature minimum up to
$z_\mathrm{max}=1850$~km) {is $\sigma \sim 0.042$ g\,cm$^{-2}$.
Therefore, according to equation~(\ref{eq:3}), the chromospheric density may increase 
by a factor of two ($\Delta\sigma=\sigma$) in $\Delta t \sim 400$~seconds}.

One hour after the flare, matter condenses again down to the 
photosphere. The magnetic field maps indicate that while the flare is taking place, most of the photospheric magnetic 
field is below detectable limits. One may speculate that the heating process has caused an increase of the gas 
pressure, thus producing an expansion of the magnetic structures and reducing their field strength. As the material 
evaporates, the gas pressure decreases gradually and the magnetic field tends to concentrate again trying to recover its 
original strength. However, the mass downflow observed during the last map may tend to increase again the gas pressure, 
leading again to an expansion of the magnetic field lines. This speculative scenario may explain at the same time the observed 
variations of the magnetic field strength and velocities during the pre-, flare, and post-flare phases.         

\acknowledgments

The Vacuum Tower Telescope is operated by the Kiepenheuer-Institute for Solar Physics in Freiburg,
Germany, at the Spanish Observatorio del Teide, Tenerife, Canary Islands. This project was supported in part by grant DE 787/3-1 of the German Science Foundation (DFG). The authors would like to thank C. Denker and E. Khomenko for carefully reading the manuscript.

\end{document}